\documentclass[acmsmall]{acmart}
\usepackage{enumitem}
\usepackage{graphicx}
\usepackage{verbatim}
\usepackage{adjustbox}
\setlist[itemize]{align=parleft,left=0pt..1em}
\AtBeginDocument{%
  \providecommand\BibTeX{{%
    \normalfont B\kern-0.5em{\scshape i\kern-0.25em b}\kern-0.8em\TeX}}}

\begin{document}

\title{Multi-Tenant Cloud FPGA: A Survey on Security}

\author{Muhammed kawser Ahmed}
\authornote{}
\email{muhammed.kawsera@ufl.edu}
\affiliation{%
  \institution{University of Florida}
  \streetaddress{}
  \city{Gainesville}
  \state{FL}
  \country{USA}
  \postcode{32603}
}
\author{Joel Mandebi }
\email{jmandebimbongue@ufl.edu}
\affiliation{%
   \institution{University of Florida}
 \streetaddress{}
 \city{Gainesville}
 \state{FL}
 \country{USA}
 \postcode{32603}
}
\author{Sujan Kumar Saha }
\email{sujansaha@ufl.edu}
\affiliation{%
   \institution{University of Florida}
 \streetaddress{}
 \city{Gainesville}
 \state{FL}
 \country{USA}
 \postcode{32603}
}

\author{Christophe Bobda }
\email{cbobda@ufl.edu}
\affiliation{%
   \institution{University of Florida}
 \streetaddress{}
 \city{Gainesville}
 \state{FL}
 \country{USA}
 \postcode{32603}
}

\renewcommand{\shortauthors}{Kawser Ahmed, et al.}

\begin{abstract}
With the exponentially increasing demand for performance and scalability in cloud applications and systems, data center architectures evolved to integrate heterogeneous computing fabrics that leverage CPUs, GPUs, and FPGAs. FPGAs differ from traditional processing platforms such as CPUs and GPUs in that they are reconfigurable at run-time, providing increased and customized performance, flexibility, and acceleration. FPGAs can perform large-scale search optimization, acceleration, and signal processing tasks compared with power, latency, and processing speed. Many public cloud provider giants, including Amazon, Huawei, Microsoft, Alibaba, etc., have already started integrating FPGA-based cloud acceleration services. While FPGAs in cloud applications enable customized acceleration with low power consumption, it also incurs new security challenges that still need to be reviewed. Allowing cloud users to reconfigure the hardware design after deployment could open the backdoors for malicious attackers, potentially putting the cloud platform at risk. Considering security risks, public cloud providers still don't offer multi-tenant FPGA services. This paper analyzes the security concerns of multi-tenant cloud FPGAs, gives a thorough description of the security problems associated with them, and discusses upcoming future challenges in this field of study.

\end{abstract}



\maketitle

\section{Introduction}
Due to increased performance, computation, and parallelism benefits over traditional accelerators such as GPUs, FPGAs are being integrated into the cloud and data center platforms. For the last few decades, technology market had an rising demand for high-speed cloud computation. Commercial cloud providers started using FPGAs in their cloud, and data centers permit tenants to implement their custom hardware accelerators on the FPGA boards over the data center. The integration of FPGAs in the cloud was followed after Microsoft has published its work on Catapult in 2014\cite{Catapult}. Since then, it has become a niche technology for cloud service platforms, and major cloud provider giants, e.g., Amazon\cite{amazon}, Alibaba\cite{alibaba}, Baidu\cite{baidu}, Tencent\cite{tencant}, etc., have integrated FPGAs into their platform. In this forecast, global FPGA market is projected to achieve 9.1 billion market space by 2026 following a compound annual growth (CAGR) of 7\% \cite{fpga_forecast}.
For computationally intensive workloads like artificial intelligence, image and video processing, signal processing, big data analytics, genomics, etc., users can exploit FPGA acceleration in cloud platforms \cite{cloud_security_1}. 
FPGAs offer unique advantages with traditional CPU and GPU in terms of computation and flexibility. We explain these features with four concrete examples. 1. Microsoft Bing search engine experienced
25 percent latency reduction and 50 percent increase in throughput in their data centers. \cite{Catapult}. 2. Using Amazon AWS FPGA F1 instance, the Edico Genome project\cite{example_gnome} has over ten times speed and performance increase for analyzing genome sequences. 3. Introduction of Xilinx Versal FPGAs for real-time video compression and encoding in the cloud platform have significantly reduced the operating cost by reducing the encoding bitrate by 20\%, \cite{example_ai}. and 4. According to this survey \cite{example_video}, for state-of-the-art neural network implementation, 10x better speed and energy efficiency was gained using FPGA accelerators in data centers.

 For maximum utilization, a single FPGA fabric has been proposed to be shared among the cloud users by exploiting and leveraging the partial re-configurable characteristics of FPGAs\cite{multi_fpga_chen}, \cite{multi_fpga_aljahdi}, \cite{multi_fpga_byma}. This crucial property of FPGA allows reconfiguring any region in a deployed environment. Also, another key property of FPGA is the ability to allow reconfiguration of a part or region in run-time conditions. This notion of sharing the FPGA fabric resources among multiple tenants or users is called as multi-tenancy. FPGA sharing/partitioning could be established in two different ways : \textit{temporally} and \textit{spatially}. In temporal sharing, the entire FPGA fabric is allocated to  users/tenants over different scheduled time slots. FPGA fabric can be simultaneously divided into multiple logically isolated regions and allocated to different tenants. This sharing technique is referred to as spatial multiplexing and is mostly investigated by academic researchers. Industrial cloud providers avoid spatial sharing due to security
 and privacy concerns.

Even though enabling FPGAs in clouds improves performance significantly, research indicates that multi-tenant FPGA platforms have multiple security risks and concerns. FPGAs allow tenants to implement custom designs on the fabric, opening multiple attacks surfaces for the attackers, unlike CPUs and GPUs.  
Most of the common threats that has been going on against cloud FPGA are quite similar to threats that has been observed in the server based cloud computing. Beside this attacks, multi tenancy can lead for some of the attacks that is related to hardware surface, such as side channel attacks \cite{sidechannel1} where sensitive information of the hardware surface is stolen by invasive/non invasive probing, or covert channels creation in which attacker can create a hidden channel between each other to transmit confidential information \cite{covert1}, \cite{multi_cross_1}. Malicious attackers can also launch Denial of Service (Dos) in any layer of the cloud system including user or infrastructure level \cite{multi_rowhammer_1}. One of the major attacks beside this also uploading corrupted or malicious bitstreams uploaded to the FPGA fabric which can lead to shutdown or faulty results \cite{bitstream1}. In hardware level structure, unlike software based attacks cloud FPGA based attacks includes mainly hardware fabric intervention and manipulation whereby the malicious contribution can lead to to short circuits fault, performance degradation or shutdown \cite{multi_voltage}. Currently, major cloud FPGA vendors such as AWS are already offering single tenant bitstream deployment. Whereas, multi-tenant cloud scope are being under active research. In context of multi-tenant cloud FPGA, the security risk is more severe and intense as the single FPGA is shared among different users which expose the hardware surface more widely. As a result, this security concerns should be the top priority of the researchers and as multi-tenant cloud FPGA in the future will expand due to immense pressure of migrating to cloud platforms. Through out the paper we tried to cover five fundamental key concepts about the multi-tenant cloud FPGA security. First, in Section \ref{Section:Background} we tried to introduce about cloud FPGA multi-tenant concepts and related topic backgrounds. Section \ref{Section:Deployment} discuss about the detailed cloud FPGA infrastructure and possible deployment models. Categorized threat models were introduced in Section \ref{Section:ThreadModels} and in later section current established and proposed attacks were described.  Implemented countermeasures and mitigation's approaches were summarized in Section \ref{Section:Countermeasures}. We also discussed some possible research scopes in context of multi-tenant cloud FPGA, which are previously implemented in FPGA SoC but not implemented yet in cloud FPGA.

\paragraph{\textbf{Organization}}

We introduced a thorough understanding of fundamental subjects in multi-tenant cloud FPGA research, such as FPGA, cloud computing, virtualization, and OS support in Section \ref{Section:Background}. We discussed the various methods for enabling FPGAs in the cloud in Section \ref{Section:Deployment}.
Section \ref{Section:Industry} provided a summary of the industrial development and current deployment models of cloud FPGA. We created a section called \ref{Section:ThreadModels} where we categorized the threat models in order to better identify the potential attack surfaces. In section \ref{Section:Attacks}, we compiled a substantial quantity of multi-tenant FPGA attacks research and illustrated their proposed building blocks of attack circuits.
The various defenses and methods against the suggested attacks were covered in the following section \ref{Section:Countermeasures}. In Section \ref{Section:Discussion}, we expressed our opinions on virtualization security risks and future challenges.

\section{Background}

\label{Section:Background}
\noindent

\subsection{FPGAs}
\label{Subsection:FPGA}
Field Programming Gate Arrays (FPGAs) are re-programmable integrated circuits that contain an array of finite programming blocks. FPGAs provides flexibility, reliability, and parallelism compared to the traditional CPU architectures. Unlike CPUs, FPGAs have massive parallelism in performing computation. In Fig. \ref{fig:fpga} a FPGA architechture block was illustrated which includes of configuration logic blocks (CLBs) surrounded with programmable switch boxes (SBs), and data input/output (I/O) blocks.  CLBs are the fundamental logic building blocks in FPGA to perform any arithmetic function in digital logic perspective. CLBs are placed in a array of bus interconnects which are mainly controlled by programmable switch-boxes. Switch-box controlled the routing of wires around CLB blocks and hence provide a re-configurable computing platform. Every CLBs contains three fundamental digital logic based blocks lookup tables (LUTS), flips-flops and multiplexer which are used for implementing arithmetical logic function or storing data. The I/O blocks ensures bidirectional data communication between FPGA board interface and connected peripherals and external devices.

\begin{figure}[h]
	\centering
	\includegraphics[width=10cm]{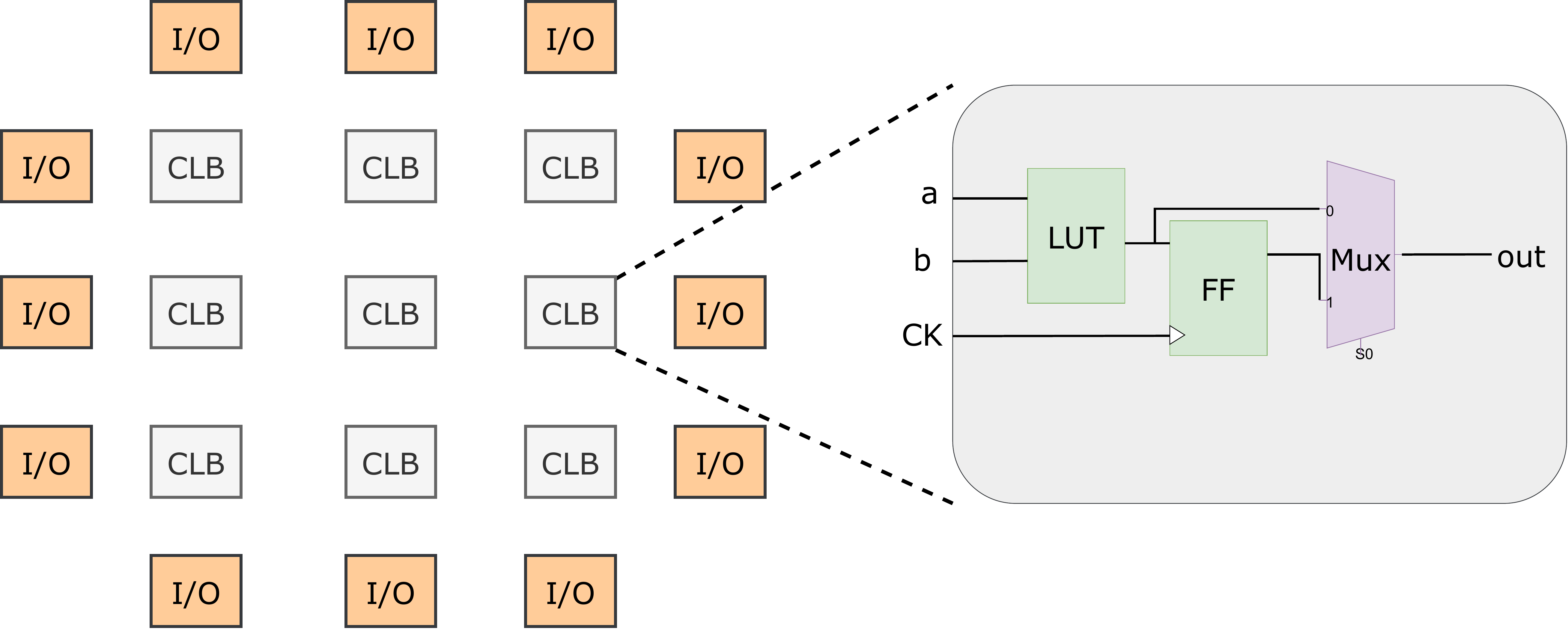}
	\caption{FPGA architecture has three fundamental digital logic components for performing boolean arithmetic functions. They are : 1) Configuration Logic Blocks(CLB), 2) Switch-box(SBx) and 3) Input/output (I/O) cells. Every individual CLB contains three major logic blocks : Look Up Table (LUT), a Mux and flip-flop (ff).}
	\Description{Cloud computing deployment and service models.}
	\label{fig:fpga}
\end{figure}

\subsection{FPGA Design Flow }
Through the first 20 years of the FPGA development period, hardware description languages(HDLs) such as VHDL and Verilog have been used and evolved to implement a circuit in FPGAs. HDL languages demand a deep understanding of underlying digital hardware. However, the improvement and development of high-level-synthesis (HLS) design tools e.g. Vivado HLS, Labview, Matlab, and C++, has added a new dimension in the FPGA design flow and can abstract graphical block or high-level  code to equivalent low-level hardware circuitry. Without achieving detailed understanding of low level hardware circuitry design, any designer can start an HLS language tool that will compile and interpret the HLS circuitry to low-level Register Transfer Level (RTL) abstraction.

Different synthesis tools are used in the FPGA design pipeline to synthesize and translate HDL codes into a netlist after they have been evaluated and simulated for behavioral accuracy and intended functionality (Xilinx, Intel Quartus, Synopsys etc.). A complete route of component connections is provided in the netlist, which comprises a description of all FPGA block connections (LUTs and registers). The components of the FPGA board are then uploaded to the FPGA board by the synthesis tool after being transformed into a bitstream binary text file. Often designers write test benches with different HDL languages, which will wrap the design code and exercise the FPGA implementation by asserting inputs and verifying outputs. Generally, in FPGA design flow, designer can simulate and test the design in three different phases: pre-synthesis, post-synthesis, and post-implementation. This whole simulation and verification phases often require more time than designing the HDL block itself. 

\begin{figure}[h]
	\centering
	\includegraphics[width=12cm]{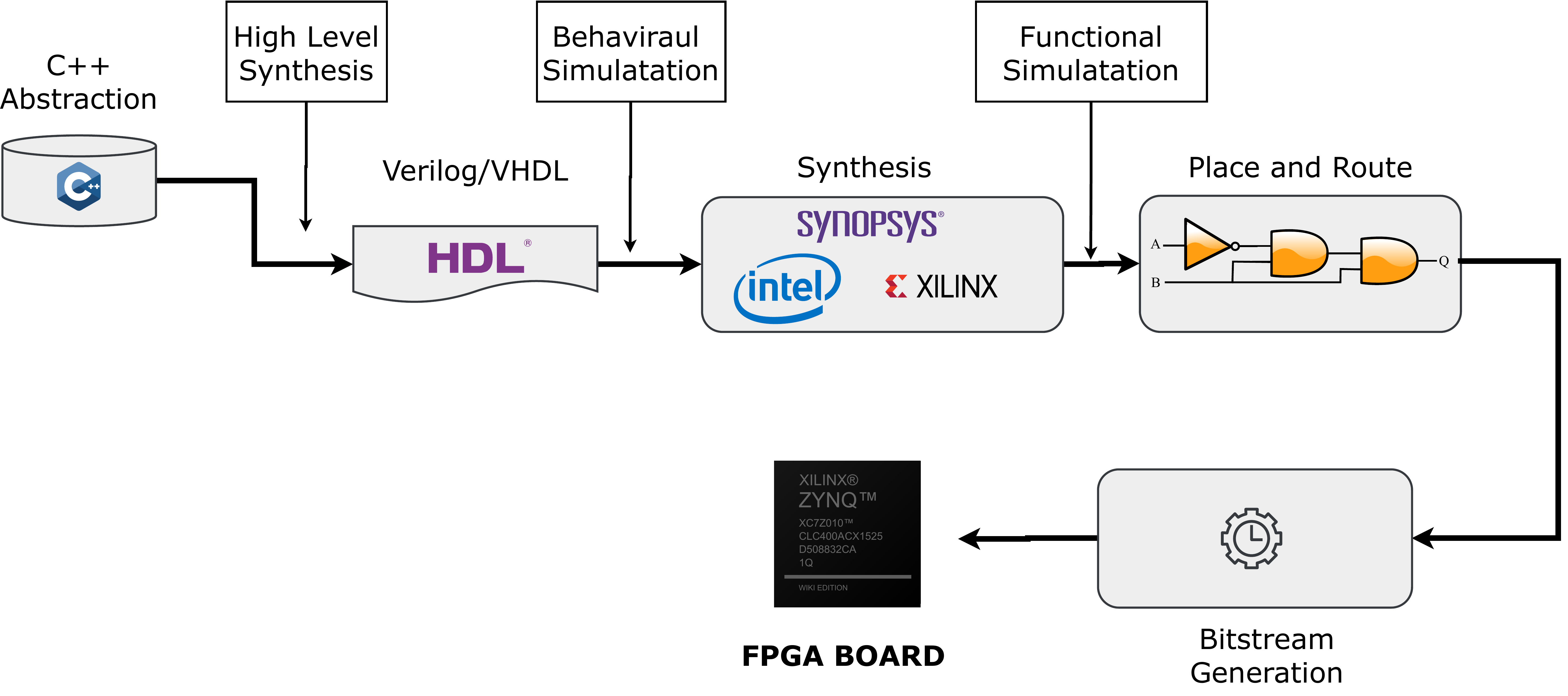}
	\caption{
	 	 FPGA design flow diagram. After high-level abstraction of desired hardware design to low-level hardware circuitry, hardware block design is simulated and synthesized using different synthesis tools (Xilinx, Intel, etc.). Next, synthesis tools generate the netlist containing the mapped design's detailed route and finally upload it on a physical FPGA board. }
	\label{fig:fpga}
\end{figure}
\subsection{Partial Reconfiguration}

The configuration and hardware logic layers are two separate levels that make up an FPGA fabric. Lookup tables (LUTs), flip flops, memory blocks, and switches listed in Section \ref{Subsection:FPGA} are all parts of the hardware logic layer. The configuration memory layer stores a binary file with all the hardware circuitry information. All configuration chores, such as recording the values of LUTs, flip-flops, and memory, managing the voltage levels of input/output pins, and routing data for interconnections, are stored in this binary file. An individual can reload the whole functionality of the FPGA board by uploading a fresh bitstream file onto this configuration memory. Bitstream transfers to FPGA are often made using JTAG or USB devices. Partial Reconfiguration (PR) enables the modification of a specific FPGA region dynamically while the rest of the FPGA region continues to run without interruption. According to the functional specifications, an FPGA fabric can be configured and partitioned into two different regions: static and dynamic. In the dynamic region, the configuration memory of the FPGA fabric is uploaded with partial bitstream without altering the functionalities of the other region, which introduce flexibility and improve resource utilization. While in the static region, the whole bitstream of the FPGA fabric is reconfigured and reuploaded, removing the previously configured bitstream.

\subsection{Cloud Computing}

The term ``\verb|Cloud|'' is coined to indicate the technology \textit{Internet} and it refers to an internet-based computing platform where various online services, including servers, storage, and websites, are supplied to the clients' PCs via the internet. Cloud computing takes advantage of a cloud network's physical resources (software, servers, and layers) being accessible everywhere. These materials are shared with cloud customers by cloud provider vendor companies, according to \cite{nist}. For particular customers, cloud services typically establish a virtual machine with their own addresses and storage. The goal of virtualization is to separate the resources (hardware and software) into various virtual servers, each of which functions as a separate, independent server.
The accessibility of virtual servers through the internet via various connectivity platforms is one of the advantages of cloud computing.

\subsection{Cloud Computing Architecture} 
\noindent
\label{cloud_models}

One of the most extensive and understandable explanations of cloud computing is defined by the  National Institute of Standards and Technology (NIST). This explanation illustrate the platform's by three service models, four deployment models, and five key attributes \cite{nist}. The five key attributes are: 1. \textbf{Self-service on demand} (These computing resources [storage, processing ability, virtuality, etc.] can be accessed and used instantly without any human interaction from cloud service providers.) 2. \textbf{Broad network access} (	These computing resources can be accessed from heterogeneous devices over the internet, such as laptops, mobiles, IoT devices, etc.)  3. \textbf{Resource pooling} (Cloud computing ensures that multiple users pool the cloud resources over the internet. This pooling is called multi-tenancy, whereas, for example, a physical internet server can initiate the hosting of several virtual servers between different users), 4. \textbf{Rapid elasticity} (Rapid elasticity creates a way of automatically requesting additional space in the cloud or any other services, leading to providing scalable provisioning. In a sense, these characteristics make cloud computing resources appear infinite.) and  5. \textbf{Measured service }(The total resource used in the cloud system can be metered using standard metrics.). According to the NIST definition, each cloud provider offers users services at various levels of abstraction, or service models. These are the most typical service models:

\begin{enumerate}[leftmargin=*]
	\item \textbf{Software as a Service (SaaS)} :	In the SaaS service model, consumers only have limited admin control of running and executing provider cloud applications and services on cloud infrastructure . Cloud applications can be executed and accessible from various client devices over the internet. The most well-liked service model is SaaS.
The most popular SaaS service products include Gmail, Slack, and Microsoft Office 365. A SaaS model could be very effective to achieve low- cost and flexible management platform. \cite{saas2} 
	whereas public cloud provided in background manages software layer executions. 
	
	\item \textbf{Platform as a Service (PaaS)}:The customer has the authority to install applications and to develop the libraries, services, and tools the provider supports under the PaaS service model. The cloud platform's fundamental foundation architecture is not managed or controlled by the user.For the development of their own applications, users can employ a variety of PaaS platforms. Examples of PaaS providers are Google App Engine, Amazon Web Services (AWS), Microsoft Azure, etc.
	\item \textbf{Infrastructure as a Service (IaaS)}:
The IaaS model is a low-level abstraction service, in contrast to the PaaS model, that gives users additional freedom for resource pooling of the deployed software's resources.  \cite{nist}. Google Cloud, Amazon Web Services and Openstack are the three biggest IaaS providers. Openstack \cite{openstack} is an open-source-based leading cloud management platform used to manage IaaS service elements (storage, networking, and I/O). OpenStack enables a cloud
		user to easily add servers, storage, and networking components to their cloud.
		Many FPGA virtualization systems are developed and implemented in cloud and data centers by
		utilizing OpenStack framework \cite{openstack2}.

\end{enumerate}

\begin{figure}[h]
	\centering
	\includegraphics[width=8cm]{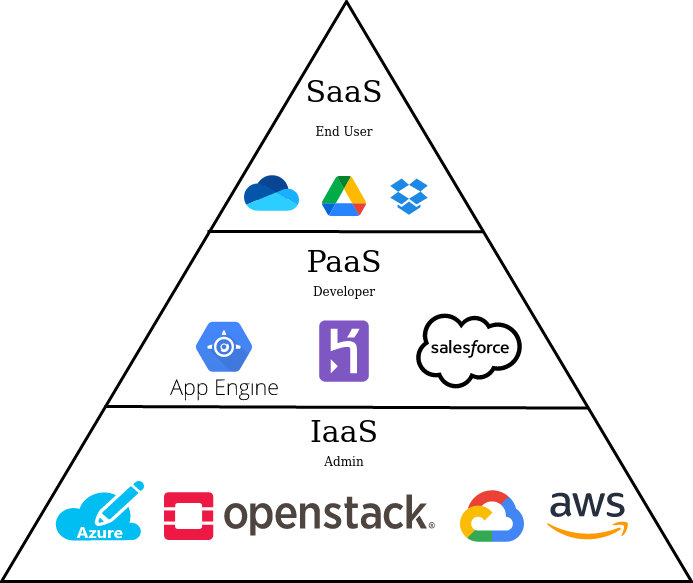}
	\caption{The main three cloud service models are: Saas, PaaS and IaaS. SaaS model allows limited admin control of running and executing on cloud infrastructure. PaaS streamlines and reduces the cost of creating and delivering apps. IaaS is a cloud service paradigm that provides full control over cloud infrastructure.  }
	\Description{Cloud computing deployment and service models.}
	\label{fig:services}
\end{figure}

\subsection{Virtualization}

Virtualization technologies are now widely used and significant in the cloud computing and big data industries.
It has produced various advantages, including flexibility, independence, isolation, and the capacity to share resources.
In general, virtualization is an effective method of dividing up computing resources among several users into virtual abstractions.
Virtual Machine Monitor (VMM) or Hypervisor is a piece of control software used in virtualization technologies that oversees resource virtualization for all of the virtual machines in the underlying hardware system.
Running various OS in a virtual platform instance called Virtual Monitor is the most prevalent example of virtualization (VM). 

The term "FPGA virtualization" describes the abstraction of FPGA resources and their distribution among numerous users or tenants.
In virtualization, an overlay architecture, also known as a hardware layer, is placed on top of the FPGA fabric.
Multi-tenancy, isolation, flexibility, scalability, and performance are the goals of FPGA virtualization.

\subsubsection{SHELL}
One of the most important parts of FPGA virtualization is defining the SHELL, and ROLE abstraction, which is described as \textbf{SHELL SHELL Architecture {SRA} } in many literature's \cite{virt_survey_1}. Inside an FPGA fabric, the shell is a static region that contains some pre-defined and pre-implemented deployments. Usually, the shell region contains three basic elements: a. Control Logic b. Memory Controller (DMA Controller) and c. Network Controller (PCIe). The primary role of the SHELL region is to provide necessary control logic and I/O details so data can be easily exchanged between the user and FPGA fabric. SHELL also provides isolation of core system components and hence provides security guarantees. The shell typically receives one-fourth of the DRAMs.
One of the four DDR4 DRAM controllers utilized in Amazon EC2 F1 instances is implemented in the SHELL region. 

\begin{figure}[h]
	\centering
	\includegraphics[width=6 cm]{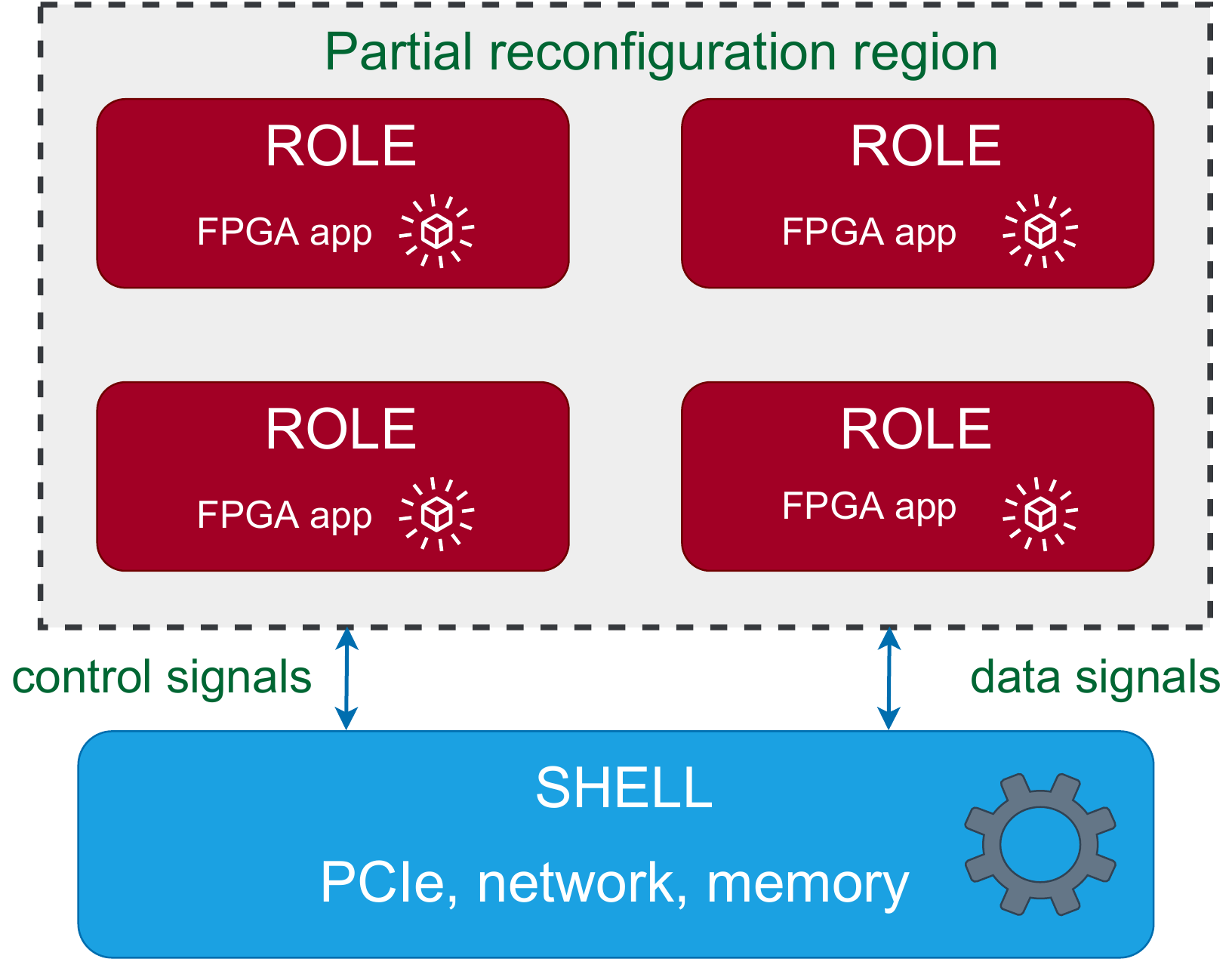}
	\caption{Architecture for SHELL and ROLE region.
The SHELL area houses the three main shared FPGA fabric building blocks (PCIe, network, and memory), which are crucial building blocks for the tenants. Shell region is isolated and secured from rest of the ROLE region. The re-configurable block known as the ROLE region houses inhabited tenants. }
	\Description{Shell Role Architecture}
	\label{fig:shell}
\end{figure}
\subsubsection{ROLE}
The ROLE section of the FPGA is called the dynamic or re-configurable region, which the user at runtime configuration can map. Bitstream mapping of the FPGA fabric can be launched either fully or partially. The role region provides elasticity and flexibility, which helps achieve higher resource utilization and improved performance.  In Fig. \ref{fig:shell} a SHELL ROLE architecture was described, indicating the isolated regions for partial reconfiguration of FPGA block. 

\subsubsection{Operating System(OS) Support}
\label{subsection:OSSupport}
For FPGA virtualization, some software applications or a OS must be deployed. Abstraction of FPGA fabric is generally available by the software stacks or a OS. Tenants can create and deploy applications into the FPGA fabric using software stacks and operating systems without having a deep understanding of the underlying FPGA hardware.
The most popular software stack for FPGA virtualization is OS since the concept of virtualization is derived from the basis of OS. However, there is currently no well-established operating system that addresses OS abstraction for reconfigurable computing systems. The literature has proposed some architecture to provide OS virtualization in cloud FPGA platforms using Linux and Windows. Besides Linux and Windows, some specialized OSs is also designed to exploit FPGA fabric's reconfigurable nature. FPGA-based cloud virtualization does not abstract core logical resources like LUTs, block memory, or flip-flops into many virtual FPGA instances, in contrast to software-based virtualization where the virtual machine monitor starts several virtual machine (VM) instances. Instead, the FPGA cloud virtualization aims to support of abstracting the FPGA hardware layer for emulating OS instructions.  Custom OSs proposed for FPGA virtualization follow the paravirtualization method, where mostly Linux kernels are modified to support FPGA hardware abstraction. FPGA based operating system are generally divided into two categories: 1) Embedded Processor OSs and 2) Re-configurable CPU . Embedded processor OSs developed for FPGA platforms normally operate on the embedded processor integrated into the same FPGA SoC. Re-configurable CPU are modified OSs that run directly on FPGA hardware fabric and emulate OS instructions even though the underlying FPGA architecture is different from traditional CPUs.  

\paragraph{\textbf{Embedded Processor OS}}
	
	HybridOS \cite{os_hybrid} is a re-configurable accelerator framework that is developed by modifying the Linux kernel and implemented in the embedded processor inside of Xilinx Virtex II SoC.   
\paragraph{\textbf{Re-configurable CPU} }

	BORPH \cite{os_borph}, modifies the Linux kernel to execute the FPGA process as a user FPGA application. The modified Linux kernel can abstract the conventional CPU instruction platform in the FPGA fabric as a hardware process. In a manner similar to a traditional processor-based system, tenants can execute system calls and perform necessary operating system functions.  
BORPH abstracts the memories and registers defined in the FPGA using the Unix operating system as a pipeline. 
	Unlike BORPH, FUSE \cite{os_fuse} leverages a modified kernel module to support FPGA accelerators in the form of tasks instead of a process. In order to decrease data transmission latency between HW and SW operations, it also makes use of a shared kernel memory paradigm. ReconOS \cite{os_recon} introduces the  multi-thread abstractions of software programs and standardized interface for integrating custom hardware accelerators. Like FUSE and BORPH, Recon uses the modified Linux kernel to develop this framework. At a high level, Feniks \cite{os_fenik}  abstracts FPGA into two distinct region: OS and multi-application regions, both of that are provided to software applications. 

To effectively connect with the local DRAM of the FPGA, the host CPU and memory, servers, and cloud services, software stacks and modules are present in the OS region.
In addition, Feniks provides resource management and FPGA allocation using centralized cloud controllers that run on host CPUs. Like an OS, LEAP \cite{os_leap} provides a uniform, abstract interface to underlying hardware resources and is similar to other proposed architectures. AmorphOS’s OS \cite{os_amorph}, divides the FPGA region into a small fixed-size zone, a.k.a morph lets, which provides the virtualization of the FPGA fabric. AmorphOS performs the sharing of the hardware tasks by either spatial sharing or the time-sharing schedule method.

\section{ Cloud FPGA Infrastructure and Deployment Model }

\label{Section:Deployment}

FPGAs have mostly been used in the verification phase of ASIC designs over the previous ten years, where the ASIC design was implemented for validation and verification phases before it was actually produced.
Additionally, specialized markets and research programs had some other applications.
However, FPGAs are gaining popularity as an alternative for CPUs and GPUs due to high performance processing and parallelism.
FPGA boards are both available on the market today as supported devices that may be connected by PCIe ports or as part of the same System-on-Chip (SoC). Recent trends indicate that the integration of  FPGA is growing exponentially in cloud platforms to provide tenants with designing and implementing their custom hardware accelerators.   

\begin{figure}[h]
	\centering
	\includegraphics[width=11 cm]{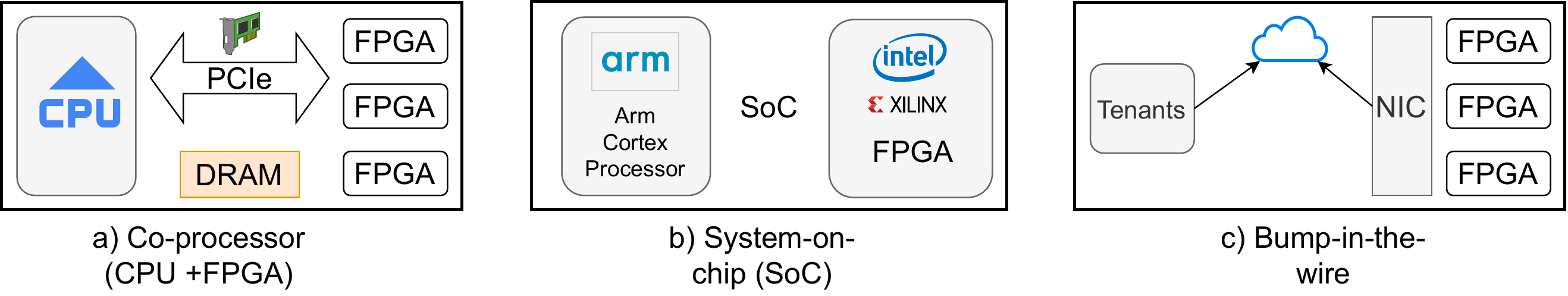}
	\caption{FPGA deployment models in the cloud. a) In the co-processor model, PCIe connections are used to connect FPGA boards to CPUs in data centers.
b) In the SoC model, the FPGA and CPU are mounted on a chip die.
c) The bump-in-the-wire concept uses FPGAs in the data centers, which tenants can access via NIC protocols. }
	\Description{FPGA deployment models in the cloud.}
	\label{fig:deploy}
\end{figure}

There are typically four basic methods for deploying FPGA boards in cloud data centers. Fig \ref{fig:deploy} shows the different FPGA deployment models on the cloud.

\subsection{Co-processor}
\label{subsection:deploy_co-processor}
In the first approach, FPGA is considered a co-processor. FPGA and CPU are located in a same node in a data center and can be accessed by the PCIe network. However, the total count FPGA boards in the data center is proportional to the total number of CPUs, and the FPGA board cannot run independently. Xilinx introduced the first CPU+FPGA integration in 2013 for embedded devices named a Zynq SoC \cite{zynq}. This SoC platform is integrated with ARM cortex processor and FPGA programming block in the same die. Removing communication ports to the CPU reduces latency and increases overall performance. The AXI-based communication protocol introduced the communication between FPGA and CPU. In year 2014, a project lead by Microsoft implemeted the idea of integrating FPGA and CPUs in a datacenter named as Catapult \cite{catapult}. In this project, Microsoft aimed at augmenting CPUs with an interconnected and configurable programming FPGA block. The project was first used as a prototype in the Bing search engine to identify and accelerate computationally expensive operations in Bing’s IndexServe engine. The experimental project was very successful and generated outstanding results while increasing a dramatic increase in the search latency, running Bing search algorithms 40 times faster than CPUs alone \cite{catapult2}. These results pushed Microsoft to extend the other web services.In 2015, Intel acquired Altera and its (Xeon+FPGA) deployment model that integrates FPGAs along with Xeon CPUs. In the backend, Intel introduces a partial reconfiguration of the bitstream, where static region of FPGA is allocated. This reconfiguration is referred to as blue bitstream as it can load the user bitstream (named as a green bitstream) into the re-configurable blue region. This interaction is handled by an interface protocol called Core Cache Interface (CCI-P).

Amazon announced its affiliation with Xilinx for accelerated cloud technology using FPGA in 2016.
This project was controlled under an AWS shell module where the user logic was configured dynamically in the FPGA hardware. Tenant are provided with amazon-designed software API shell to avoid any security damanges. In the last recent years, Baidu \cite{baidu}, Huawei \cite{huawei}, Tencent \cite{tencant}, and Alibaba also started the recent trends of integrating FPGA and CPU.

\subsection{Discrete}
\label{subsection:deploy_discrete}

FPGA board can be also deployed independently as an individual separate component which eliminates the ncessity of deploying CPU along with FPGA boards. This discrete approach considers deploying the FPGA board as separate standalone component. This setup is independent from the CPU and FPGA board is directly connected to the network. For example, NARC \cite{narc} is a standalone FPGA board which is connected through the network and capable of performing high computational workloads and network tasks. By using OpenStack platform, Smart Applications on Virtualized Infras-tructure (SAVI) project  \cite{deploy_toronto} deployed a cluster of discrete FPGA boards that can communicate privately in a closed network. 
IBM announced cloudFPGA project of accommodating of  1024 FPGA boards in total to a data-centre rack, in 2017 \cite{ibm}. IBM deploys these FPGA racks as stand-alone resources for hardware and avoids the common way of coupling FPGA boards with CPUs in data-centres. IBM  standalone setup has shown a increase 40x and 5x in latency and output paramater  \cite{ibm}.

\subsection{Bump-in-the-wire} 
\label{subsection:deploy_wire}

Bump-in-the-wire model refer to a setup where FPGAs are placed on a server between the
Network Interface Card(NIC) and the network. This allow FPGAs to communicate directly over the network and process data packets receiving from different users through the internet.  Bump-in-the-wire architectures experienced a dramatic
reduction in latency as layers between the communication path are reduced. Exposing the FPGA resources over the network have unique benefits of providing offload computation without interacting with CPUs and GPUs. Users can directly send packets
to the servers which is processed by the routers and later forwarded to the destined FPGA board using software defined network(SDN) protcols. 
The famous Microsoft Catapult has followed the bump-in-the-wire deployment model to connect their FPGA boards in the top-of-the-rack switch (TOR) rack.

\subsection{System-on-chip (SoC)}

\label{subsection:deploy_soc}

SoC FPGA devices integrate microprocessors with FPGA fabric into a single board. Consequently, they
provide higher reliability and reduced latency and power consumption between the
processor and FPGA. They also include a rich set of I/O devices, on-chip memory blocks, logic arrays, DSP blocks, and
high-speed transceivers. Currently, there are three families of SoC FPGAs available on the market by Intel, Xilinx, and Microsemi \cite{Jin2020}. The processors used in the FPGA SoC have fully dedicated “hardened” processor blocks. All three FPGA vendors integrated the full-featured ARM® processor and a memory hierarchy internally connected with the FPGA region. Integrating ARM processors and FPGA blocks on the same piece of a silicon die significantly reduces the board
space and fabrication cost. Communication between the two regions consumes
substantially less power than using two isolated devices.

\section{Industrial Evolution of Public Cloud FPGA Accelerators}
\label{Section:Industry}
\begin{figure}[h]
	\centering
	\includegraphics[width=8cm]{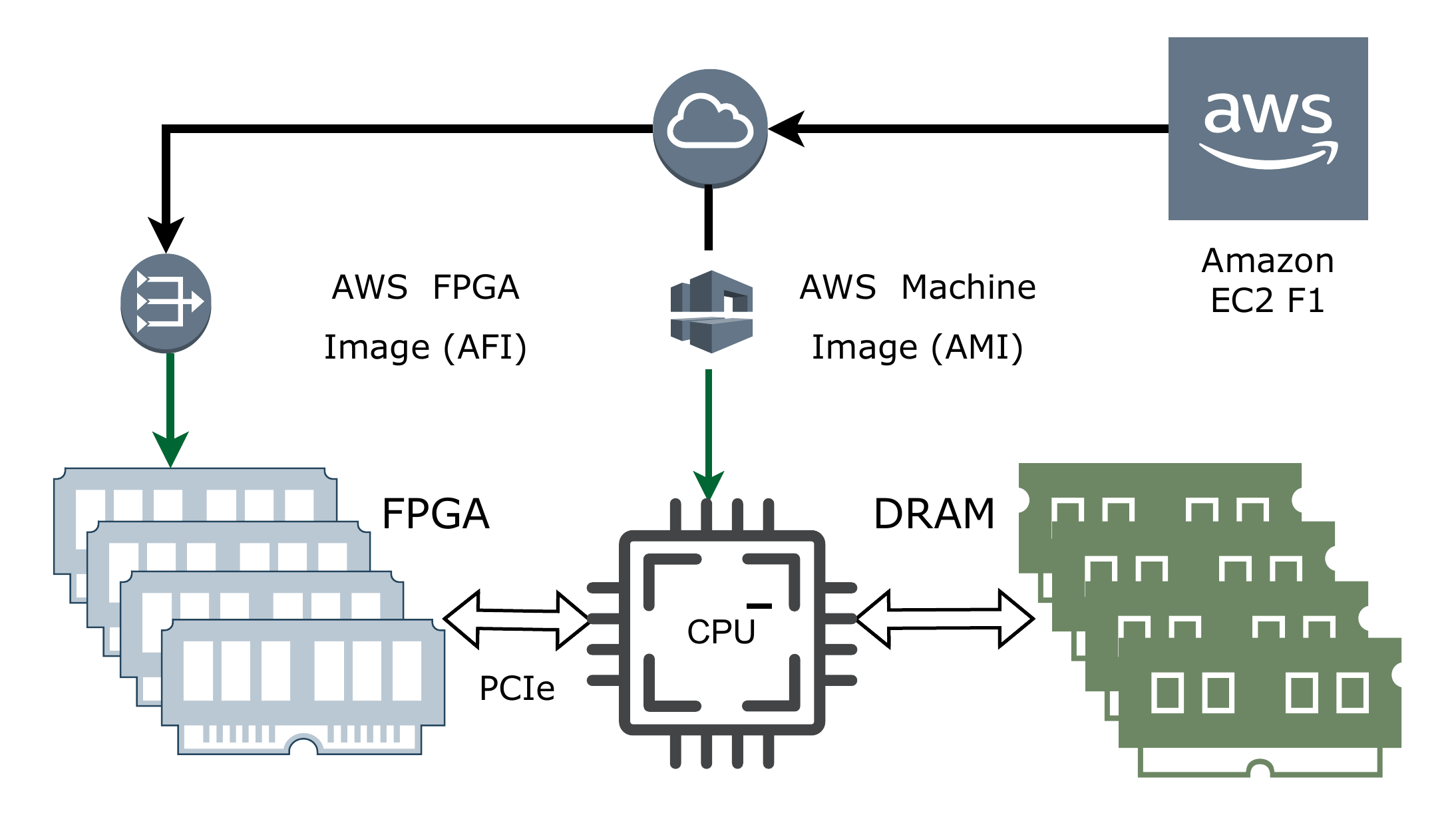}
	\caption{Amazon AWS EC2 F1 instance architecture. After bitstream checking , generated netlist (AWS FPGA Image) is uploaded in the FPGAs and respective software program (Amazon Machine Image) is uploaded in the CPU.  }
	\Description{Amazon AWS EC2 F1 instance architecture }
	\label{fig:aws}
	\end{figure}

In recent years, public cloud providers offering FPGA boards to users/tenants in their 
data centers. Users or tenants normally go for pay-per-use to access FPGA resources, and the control
is assigned to users for a specific time slot. Tenants can speed up
their application performance by designing custom hardware accelerators and implementing
them in their assigned region. Amazon AWS\cite{amazon}, Huawei Cloud\cite{huawei}, and Alibaba Cloud  \cite{alibaba} has started offering rent of the Xilinx Virtex Ultrascale+ architecture in their cloud platform. Xilinx Kintex Ultrascale is offered by Baidu \cite{baidu} and Tencent Cloud \cite{tencant}. Alibaba and OVH are currently offering Intel Arria 10 FPGA boards on their cloud. Xilinx Alveo accelerators are available through the Nimbix cloud platform \cite{Nimbix}. Normally, a cloud providers host several FPGA boards on their servers. Hypervisor of the server assigned each FPGA board to a single tenant.\\
Still now, the offerings from the major cloud providers can be divided into three main properties : 

\begin{itemize} [leftmargin=*]
	\item 	Most providers would provide user permission to upload any synthesized netlist into the FPGA
	board, including some limitations and design rule checks.
	\item  Some providers would be given the advantage of working in a high-level synthesis (HLS) platform
	instead of uploading RTL netlist.
	\item  All public cloud providers offering FPGA cloud accelerations services follow the
	co-processor deployment model (Section \ref{Section:Deployment}) in their data center servers. Bump-in-the-wire
	and System-on-chip(SoC) is not available today in public clouds.

\end{itemize}
Because most cloud providers today have similar architecture and naming conventions to AWS, we have dedicated a subsection for the AWS EC2 F1 platform. Fig \ref{fig:aws} shows the AWS EC2 F1 architecture and deployment model.

\subsection{Amazon Web Services (AWS) EC2 F1 }

Since 2016, In order to provide single-tenant FPGA cloud access, AWS has continued to offer FPGA-accelerated multi-tenant cloud services. These FPGA-based cloud services, also known as AWS F1 service, includes three model instances : f1.2xlarge, f1.4xlarge, and f1.16xlarge. This naming convention represents twice the number of FPGA instances present in the module. For example, f1.2xlarge refers to 1 FPGA board, 4xlarge refers to 2 FPGA board instances, and 16xlarge refers to total 8 FPGA boards. Every virtual machine model instance contains 8 CPU cores (virtual), 122 Gib of memory DRAM, and storage limit of 470GB(SSD). Four DDR4 DRAM chips that each FPGA board can directly access make up the AWS F1 instance architecture.
Each FPGA instance board may get up to 64GiB of RAM in total. A PCI Gen 3 bus connects the FPGA card to the server. Users cannot upload their unique hardware design directly to the FPGA in AWS. AWS uses the tools from Xilinx to create bitstreams in its own way. After completing all Design Rule Checks (DRCs), AWS generates the final bitstream (Amazon FPGA Image - AFI) \cite{amazon}.  

\section{Threat Models}

\label{Section:ThreadModels}
 We divide the threat models into four categories in order to better comprehend the potential risks to cloud FPGA users : (1) Intra tenant adversaries - Trojans (2) Inter co-tenants adversaries, and (3) Inter-node adversaries and (4) Malicious network traffics.

	\subsection{ Intra Tenant Adversaries (Trojans) }

Modern hardware design flow involves multiple designs and verification stages. Even in the design
stages, many combinations of software and hardware blocks (referred to as Intellectual IP") consist of
different levels of the highly optimized environment. Hardware IPs are heavily customized
in various sections according to the necessity of fields like Signal Processing, Video, and Image
Encoding, Quantum and Artificial Intelligence Computations, etc. Also, no single companies are
responsible for developing the whole complex IP components. Rather, hundreds of third-
IP designer companies and IP Integrators are involved in the entire FPGA hardware design
process. This complex design and integration process opens a door for several security attacks on a target FPGA fabric \cite{tehranipoor_trojan}. In hardware design trends, designer tenants use existing third-party IPs
by providing licensing fees to cut the cost of designing complex IPs. Often non-trusted third-party
IP cores or EDA tools are integrated into different stages of FPGA design life and are susceptible to
numerous attacks, such as HT injection, IP piracy, cloning, and tampering \cite{trojan_2}.

	\subsection{ Inter Tenants (co-tenant) Adversaries }

In a multi-tenant spatial sharing model, users can have a stake in a large computation resource pool, leading to numerous resource-sharing threats. Even though the allocated tenants are logically isolated, they are located in the same hardware fabric and share the same resources. These vulnerable resources include the shared DRAM, PCIe, and Power Distribution Network (PDN). A malicious adversary can exploit these advantages to launch a variety of attacks. In research \cite{multi_cross_1}, \cite{multi_cross_2}, the cross-talk and covert channel communication has been extensively explored between logically isolated co-tenants. A large power-hungry adversary tenant could potentially pull over the entire FPGA board in extreme cases.  \cite{survey_bobda}. A malicious co-tenant could also cause temperature and voltage fluctuations and affect other users by introducing row hammer attacks on shared DRAM \cite{multi_rowhammer_1}. 

\subsection{Inter Nodes (co-nodes) Adversaries}
Since FPGAs have shared DRAM access to multiple tenants, adversaries could introduce Rowhammer-style attacks for co-located node FPGAs. A malicious tenant could attack the shell region of the co-located nodes and launch a denial-of-service attack which could potentially block the host server. In co-located nodes, it is also possible to extract the FPGA process variation parameters by accessing shared DRAM \cite{security_fingerprint}. In this work, a dedicated physical unclonable function (PUF) circuit created a fingerprint of the shared DRAM instance. As DRAM is normally static in the data centers, acquiring the fingerprint of the shared DRAM module is equivalent to fingerprinting the attached FPGAs in co-located nodes. Recent work \cite{security_node} shows that the fingerprint of co-located node FPGAs could also be achieved by PCIe bus contention. 

\subsection{Malicious Network Traffic}
While FPGAs are connected to the cloud ecosystem through public network interfaces, they face unique security challenges as they process network packets directly from the users. Bump-in-the-wire (or Smart-NIC) deployment models have multiple risks for exposing the FPGAs over the public network. A malicious adversary can easily flood the inbound/outbound packets in data-center routers.  In research \cite{security_noc}, a timing-based side-channel attack was introduced in an NoC platform to extract the keys in the AES  algorithm. Using a  Multiprocessor System-on-Chip (MPSoC) platform, a Denial-of-service (Dos) based attack was carried out in a Internet-of-thing(IoT) connected network devices. \cite{security_noc_dos}. In the proposed attack, while the tenant application has been executing on NoC coupled IoT devices, the attacker could flood the network packets until Packet Injection Rate (PIR) crosses the threshold voltage and cause a complete shutdown.

\section{Multi-tenant Cloud FPGA Attacks }
\label{Section:Attacks}
Because cloud FPGA are installed in cloud servers and data centers, physical attacks on the protected servers and datacenters are rare.
Multi-tenant cloud FPGA assaults will often include non-intrusive attacks.
Three main types of attacks on multi-tenant FPGAs are as follows: 1. Extraction attacks 2. Fault Injection attacks 3. Denial-of-Service (DoS) attacks.

 \subsection{Extraction Attacks} 
\subsubsection{Remote Side Channel attacks}
Side channel attacks in FPGA have been explored extensively by the security researchers. Traditionally, side-channel attacks is carried out by probing voltage or electromagnetic waves and analysing collected traces by the devices either using Simple Power Analysis method (SPA) \cite{Ramesh20} or Differential Power Analysis method (DPA \cite{sidechannel1}). Instead of accessing actual physical FPGA hardware, a malicious attacker can exploit sensitive data and information by analysing the dynamic power consumption caused by voltage fluctuations of a remote FPGA board. This power consumption information  could potentially leaks the secret keys of the IPs core and design blocks. Generally, in terms of multi-tenant platform the attacker doesn't have the access of the actual physical hardware. Which nullify the all traditional electromagnetic based side channel attacks on FPGA board. Eventually, attackers find out a way of accessing the FPGA fabric power consumption data which is carried out in a remote network connection.In order for the malicious opponents to gain access to some of the LUTs in the remote FPGA, it is assumed that the attacker and victim are situated within the same FPGA fabric. In general, building delay sensors like Ring Oscillator (RO) based sensors \cite{multi_power_ro_1}  or Time-to-Digital Converters (TDC) based sensors \cite{multi_power_1} can be used to evaluate the voltage fluctuation on the remote FPGA board.   .

\begin{figure}[h]
	\centering
	\includegraphics[width=11cm]{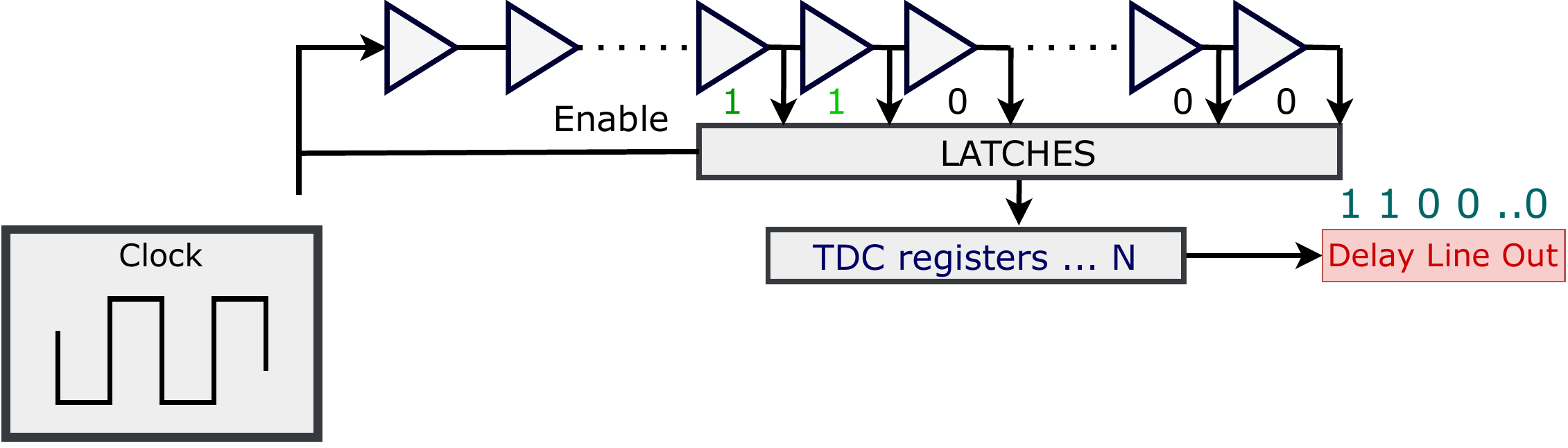}
	\caption{An example of a TDC sensor that measures delay using a series of buffers and latches. \cite{multi_power_1}.}
	\Description{Cloud computing deployment and service models.}
	\label{fig:multi_tdc}
\end{figure}

\paragraph{\textbf{Time-to-Digital (TDC) based Delay sensors}}

In FPGA fabric,  power distribution network(PDN) controls all the supply voltage distribution of all the components. PDN resources are made accessible to all users when FPGA accelerators are shared in the cloud. Every modern IC is powered by a complicated PDN, a type of power mesh network, made up of resistive, capacitive, and inductive components. Any voltage drop in these PDN can cause a dynamic change in a current. Also, variation in operating conditions will alter the supply current and voltage and hence change the PDN. By inserting a suitable sensor (Hardware Trojan) this voltage fluctuation can be read and use to launch malicious remote side channel attacks. This hardware trojan can be inserted in both pre-fabrication and post-fabrication stages of the IC development process chain. The architecture of this trojan consist a chain of delay signals which propagates through a chain of buffers. The delay signals could be designed as a Time-to-Digital Converters(TDCs), as shown in Fig. \ref{fig:multi_tdc}. In the work \cite{multi_power_1}, it is shown that the potential hardware Trojan could sufficiently sense the PDN variation events and could efficiently extract sensitive information's based on the voltage drop fluctuations. AES core running at 24 Mhz on a Xilinx Spartan-6 FPGA was used as the victim in the proof-of-concept attack described in \cite{multi_power_1} as a way to illustrate this scenario. In two alternative scenarios, the experiment was run: in one, the sensor was positioned close to the victim AES logic, while in the other, it was placed farther from the AES core. In all scenarios, the attacker is able to obtain the AES key. 

In the research, \cite{multi_power_amazon} Glamocanin et al. was successful in mounting a remote power side-channel attacks on a cryptographic accelerator running in  AWS EC2 F1 instances. The attack circuit used in this work is a TDC based delay sensor depicted in Fig \ref{fig:multi_tdc}. As a case study, AES-128 core was used and the attack could successfully recover all the secret keys of all 16 bytes from a AES-128 IP core. Although, currently in the industry deployments FPGA is mounted as a data accelerator connected to CPU via PCI e bus connections. It is expected that in coming years, FPGA + CPU integration could take the lead in the server deployments. While FPGA + CPU integration in a same system-on-chip(SoC) (Subsection \ref{subsection:deploy_soc}) dies could potentially delivers a lots of benefits in resource sharing and latency, it could have also create security bottlenecks. In future years, the rapid adaption of heterogeneous computing will probably force to fabricate more data-centers and cloud providers in one single SoC die. In the research, \cite{multi_power_soc}, Zhao et al. was successful to launch a remote base power side channel attacks from FPGA fabric in a system-on-chip deployment model (Subsection \ref{subsection:deploy_soc}). The attack was carried out using  a Ring Oscillator (RO) circuit placed in a FPGA programming logic block which can retrieve all the secret keys from a RSA crypto module placed in a CPU processor.This attack should be considered as a big security threat in perspective of multi-tenant cloud security and necessary steps should be taken to overcome this vulnerability risk.   

\paragraph{\textbf{Ring Oscillator (RO) based Delay sensors}}

\begin{figure}[h]
	\centering
	\includegraphics[width=11cm]{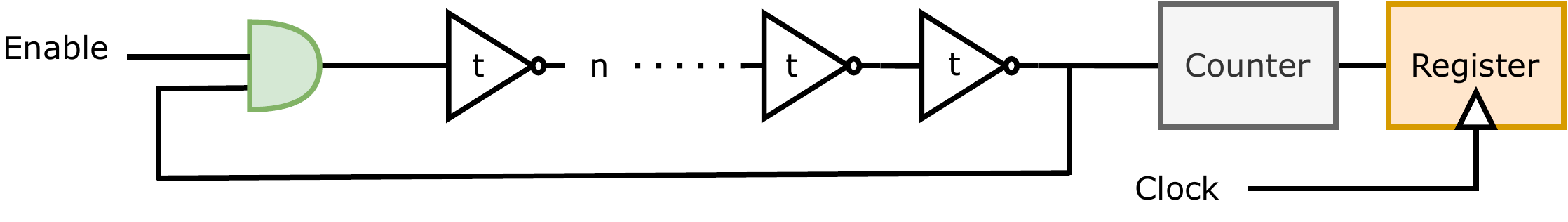}
	\caption{Proposed \cite{multi_power_ro_1} Ring Oscillator (RO) based delay sensor block for launching remote side channel attack on AES encryption core. Cascaded inverters and counters are replaced with sequential NAND gates and Johnson Ring Counter.  }
	\Description{Ring Oscillator based delay sensor}
	\label{fig:RO_basic}
\end{figure}

By measuring the oscillation frequency $f_{RO}$ of an RO-based delay sensor circuit, an attacker can extract the power supply consumption information. By using an odd number of cascaded inverters, where the output of the final inverter is sent back to the first, it is possible to create a RO-based delay circuit. Fig \ref{fig:RO_basic} depicts the traditional combinational loop-based RO circuit. Oscillation frequency $f_{RO}$ depends on the total inverters $n$ cascaded in the ring followed by this equation $f_{RO}=\frac{1}{2t_{p}n}$. Designers often integrate a sequential counter to measure oscillation ticks. Gravellier et al. proposed a high-speed on-chip RO-based sensor circuit for establishing remote side channel attacks on the FPGAs \cite{multi_power_ro_1} which can measure runtime voltage fluctuations. Using the CPA analysis technique, the proposed work successfully extracted the AES encryption core keys operating at 50 Mhz. This alternative high-speed RO circuit offers better spatial coverage and reduced overhead compared to TDC-based sensors. The proposed circuit replaced the cascaded inverters and counters with sequential NAND gates and Johnson Ring Counter (JRC).     

\subsubsection{Crosstalk Attacks}
\label{crosstalk}
In FPGA configuration block, cross talk of routing wires has been established by the research \cite{multi_cross_1}. The article demonstrates how the proximity of other long wires affects the delay of long lines (logical 1 vs 0). This delay could potentially leaks the information of the long wires and creates a communication channel even though they are not physically connected. This information leakage could create security vulnerability in a SoC platform, as often they are integrated by different multiple untrusted third parties IP. A malicious attacker could easily establish a hidden covert channel communication between the distinct cores in multi user setup. This could more severe on public cloud infrastructures where as FPGAs and CPUs are being incorporated.

\begin{figure}[h]
	\centering
	\includegraphics[width=10cm]{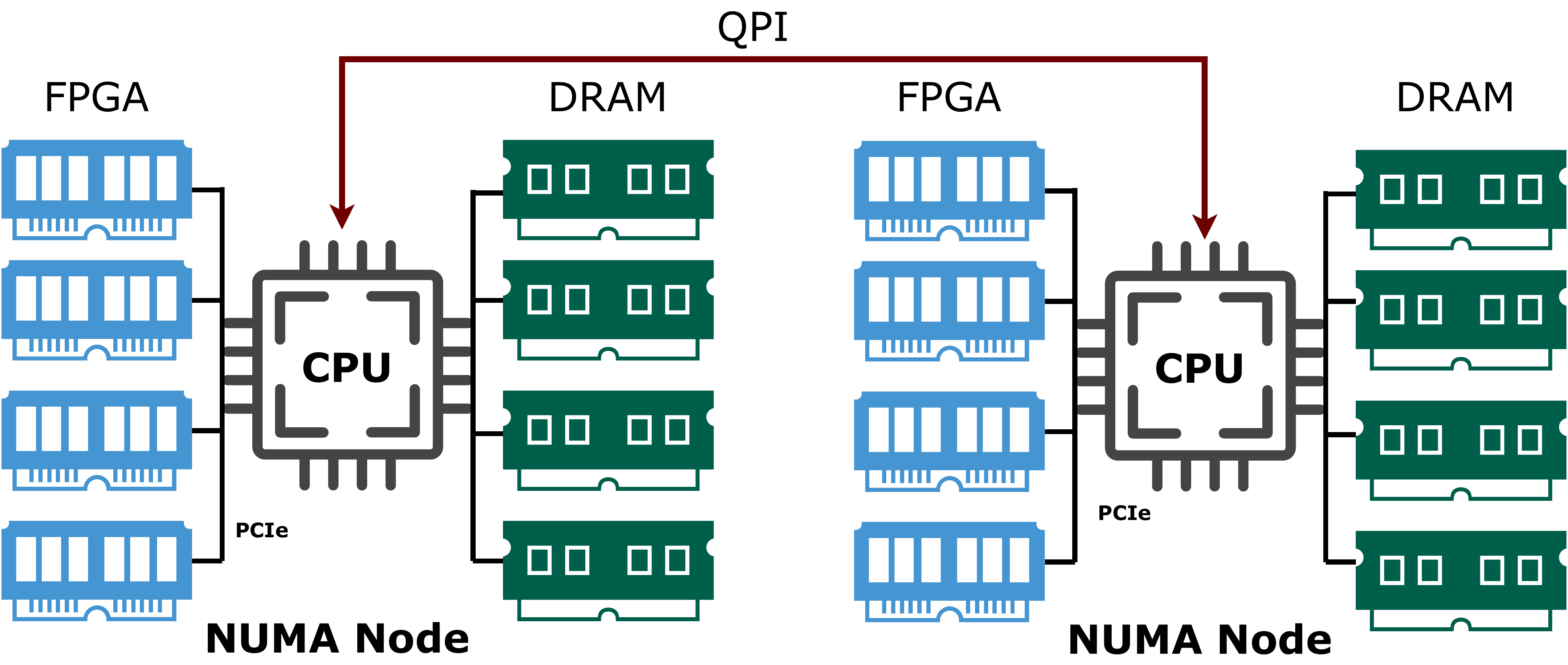}
	\caption{Single AWS F1 server , 8 FPGAs are divided between two NUMA nodes.}
	\Description {Single AWS F1 server , 8 FPGAs are divided between two NUMA nodes.}
	\label{fig:numa}
	
\end{figure}

In the paper \cite{multi_cross_2}, which is an extension of the prior work \cite{multi_cross_1} demonstrated the first covert channel between the separate virtual machine with cloud FPGAs, with reaching data transmission close to 20bps along with 99\% accuracy.The covert channel between the FPGAs was exploited by using the PCI data contentions and which opens the doors for inferring the information from the VM instances when they are co located. The authors assert that particular circuits such as Ring Oscillators (RO) or Time-to-Digital Converters (TDCs), which are typically subject to Amazon design rule checks, are not required for the attacks (DRC). Instead, it compares the bandwidth of one data transmission with a predetermined threshold value and assigns a PCI stress when the logic is reached. Three separate Amazon F1 \verb|f1.16x large| big instances (totaling 24 FPGAs) were rented for the 24-hour trial for this study. Today's cloud service providers, like AWS, only enable "single-tenant" access to FPGAs due to security risks. In these settings, each FPGA is only assigned and dedicated to the subscribing user for a predetermined amount of time, after which it may be transferred to another user. In the AWS F1 cloud instance design, each FPGA board is connected to a server by the x16 PCIe bus, and each server has a total of 8 FPGA instances. Eight FPGA instances are uniformly distributed across two Non-Uniform Memory Access (NUMA) nodes, according to a recent study using publicly available data on AWS F1 instances \cite{multi_cross_1} shown in Fig. \ref{fig:numa}.  Four FPGAs are connected as PCIe devices to each NUMA node's Intel Xeon CPUs through an Intel QuickPath Interconnect (QPI) interface. Any FPGA user on the AWS platform can perform DMA transfers from the FPGA to the server, and many designs with a little amount of logic overhead have the potential to fill the shared PCIe bus. However, users cannot directly manage PCIe transactions.
The co-location of FPGA instances will be made visible due to bus contention and interference caused by this saturation.
This concept is used to make sure that co-located NUMA FPGA instances are detected.
The two parties must make sure they are co-located in the same NUMA node prior to the attack.
They send out multiple handshake messages and wait for confirmation of the handshake response to confirm this.
The co-located FPGA instances are identified and used for upcoming communication after the answer. 
The attack is launched with two FPGAs at a time, a memory stressor and a memory tester \cite{multi_cross_2}.
The stressor constantly transmit 1 bit in the PCIe bus and the tester keeps measuring its own bandwidth during this whole transmission period. The receiver can classifies a specific hit or miss state by comparing the bandwidth with threshold value. The results shows that, it can create a fast covert channel between any two FPGAs in either direction: at 200 bps with the accuracy of the channel is 100\%. Two FPGAs are used in the attack at once, together with the memory stressor and memory tester \cite{multi_cross_2}. Throughout the whole transmission period, the stressor continuously transmits 1 bit over the PCIe bus, and the tester continuously monitors its own bandwidth.
By comparing the bandwidth with the threshold value, the receiver may identify a specific hit or miss state.
The results demonstrate its ability to establish a quick covert channel between any two FPGAs in either direction at 200 bps with 100

\subsection{Fault Injection}

A malevolent tenant could purposefully inject a fault into the FPGA during a fault-injection assault to jeopardize its security.
This could lead to a denial-of-service attack, the extraction of cryptographic secret keys, unauthorized authentication, or the facilitation of secret leaking within the system.
Attackers can employ a variety of physical tools and equipment to conduct fault-injection attacks, which can be non-invasive (such as clock glitching or voltage glitching), semi-invasive (such as local heating or laser), or intrusive (such as focussed ion beam) \cite{cloud_security_1}. In perspective of cloud FPGA, only non-invasive attacks can be carried out and inject faults by clock or voltage glitching. \cite{security_fault_majzoobi}

\subsubsection{Voltage Attacks}

\begin{figure}[h]
	\centering
	\includegraphics[width=9cm]{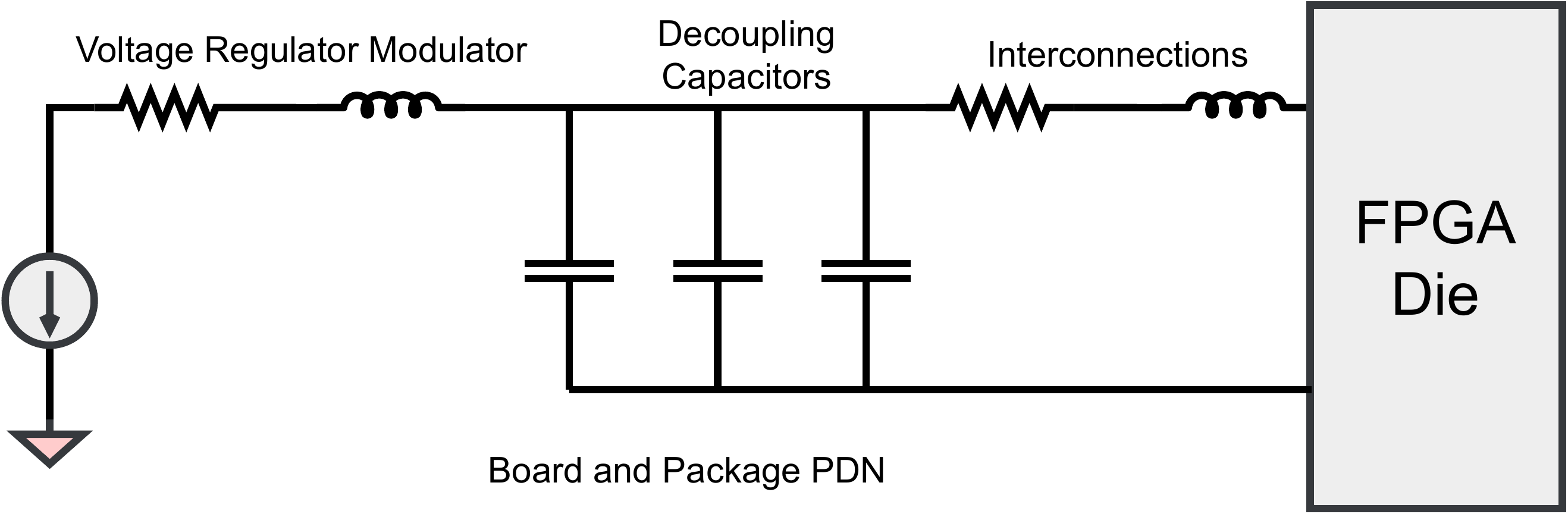}
	\caption{RC equivalent network of Power Distribution Network (PDN) block \cite{multi_voltage}.}
	\Description{RC equivalent network of PDN block.}
	\label{fig:PDN}
\end{figure}

In the paper, \cite{multi_voltage}, In a multi-tenant platform, the security of deep learning (DL) accelerators was assessed against voltage-based attacks.
The Asynchronous Ring Oscillator (RO) circuits have been used extensively in earlier research on voltage attacks.
However, Amazon AWS EC2 already employs a protection mechanism against it because this technique has been extensively researched in academic circles.
A fault injection attack is another name for a voltage attack.
An adversary circuit with hostile intentions might implement circuitry that consumes a lot of current and causes voltage drops in the chip's power distribution network (PDN). A malicious tenant could induce timing violations and potentially disrupt the default functionality of the circuit. Fig \ref{fig:PDN} shows a simplified diagram indicating the main components in a PDN block. An RC equivalent network can be used to represent an on-chip FPGA PDN circuit. In this circuit, an onboard voltage regulator controls the board voltage level into the die voltage level.
Decoupling and parallel capacitors are also present, and they aid in filtering out undesirable voltage noise. The voltage drop of this PDN block can be written by equation \ref{vdrop}. 
\begin{equation}
V_{drop}(s) = I(s) Z _{PDN}(s) \label{vdrop}
\end{equation} 
where $Z_{PDN}(s)$ is the total impedance of the PDN block in the frequency domain. In a steady-state condition, the resistive component of $Z_{PDN}(s)$ e.g. \textbf{IR} is responsible for the \textit{steady state} drop a.k.a transient drop. An effective malicious circuit could exponentially raise the transient drop (current drawn) and potentially launch a large voltage drop. As a result, it could affect the voltage propagation time in a neighboring circuit in a multi-tenant platform. Using the vendor gate clocking IP, the author introduces two novel adversarial circuits that can cause huge voltage drops in an FPGA fabric. Even though the attacker circuit is physically isolated, it can effectively launch integrity attacks on the DL accelerator running an ImageNet classification on the runtime scenario  \cite{multi_voltage}. These results indicate that current multi-tenant FPGAs are vulnerable to voltage attacks and need additional research to overcome these attacks.

\begin{figure}[h]
	\centering
	\includegraphics[width=6cm]{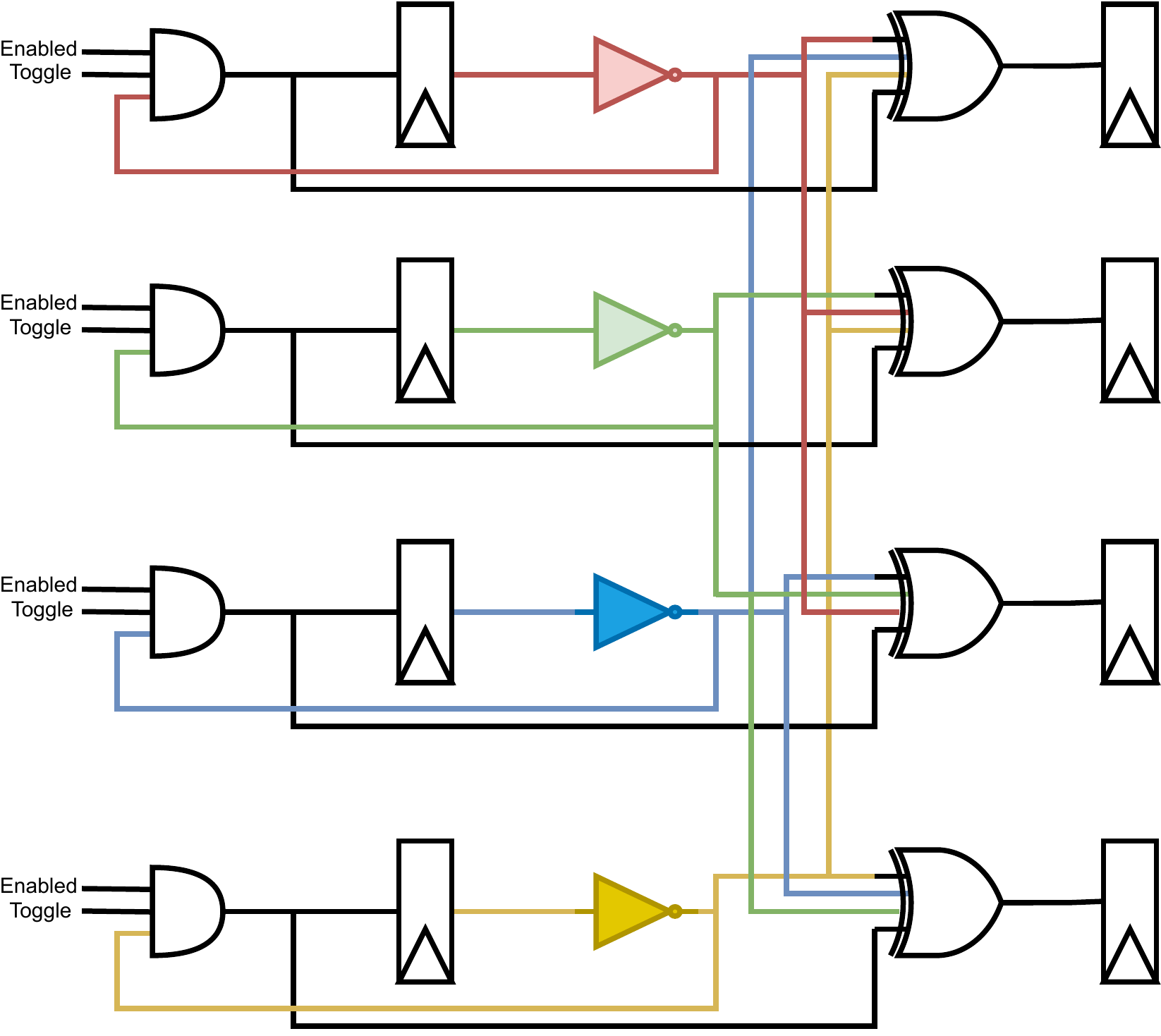}
	\caption{ Voltage attack circuit using clock-gated garbled XORs.
Adversary circuit performs clock gating, switches frequency quickly, and causes voltage drop throughout the circuit using the Intel clock control IP. }
	\Description{Clock-gated garbled XORs.}
	\label{fig:voltage}
\end{figure}

Four 4-input XOR gates make up the first attacker circuit (Fig \ref{fig:voltage}). Four toggle registers and one delayed output, which operate with a high switching frequency, serve as the inputs for XOR gates.
The toggle rate of all-gated FFs changes abruptly when a clock is enabled.
The concept behind the assaults is that voltage dips and timing errors could be introduced into the circuits if any gated clock circuitry is unexpectedly modified over a brief period. In the implementation, the adversary circuit performs clock gating by frequently switching at a high frequency while using the Intel clock control IP. This causes voltage decreases along the circuit. The final attacker circuit makes use of the attack and broadens it to include all digital signal processing (DSP) and block memory (BRAMs). The attacker circuits continue writing some patterns sequentially to every address of all BRAMs while the FPGA is working on a DL accelerator task. DSP blocks are also seeded concurrently with random initial inputs. The findings indicate that it could lead to significant timing violations in the circuits, and eventually, DL accelerators would see a significant performance hit as a result of this attack.

Alam et al. \cite{security_fault_alam}, shows that concurrent write operations in same address on dual-port RAMs could cause severe voltage drop and temperature increase inside the chip. Simultaneous write operations to the same will introduce a memory collusion leading to transient short circuit. 


%
%
 
 \subsubsection{Ring Oscillator based fault injection attacks}
 Ring oscillators are a lucrative choice to launch fault injection attacks as they achieve very high switching frequencies with minimum design and effort. 
 Mahmoud et al. \cite{security_fault_mahmoud} leveraged a ring oscillator based power hungry circuit to induce timing failures in a true random generator cryto core used for generating secret keys.  Krautter et al.\cite{security_fault_krautter} proposed a high frequency based controlled RO circuit, feasible of recovering  encrypted message in the 9th round of AES computation stages by inducing timing delays. 

\subsection{Denial of Service attacks (DoS)}

\paragraph{DRAM attacks(Row Hammers)}
In modern heterogeneous FPGA system, each FPGA board is occupied with his own attached DRAM  
which can be shared through the SoC/Datacenter. Here, the attached DRAM can be accessed from the FPGA in a unique privilege without any further monitoring. This allows one user to take a attempt of modify the DRAM of another user. This attempt, can be undetected from the CPU side and a malicious user can access the shared DRAM and launch a row-hammer attack \cite{multi_rowhammer_1}.  This research shows that, a malicious FPGA can perform twice as a typical Row-hammer in the CPU on the same system and could potentially flip four times bits in compare with the CPU attack. Further, it is much more difficult to detect on the FPGA side as user has the direct access of the FPGA's memory access operations without any intervention of CPU. 
In academia, row-hammer attacks refer to the new way of attacking shared DRAM and manipulating the user data \cite{multi_rowhammer_2}. In this method, by continuous read operation it could use the unwanted electrical charges to corrupt the data in a same row in RAM. 
Zane Weissman et al. launch a classic CPU-based Row-hammer attack along with a Jack Hammer attack(FPGA) against the WolfCrypt RSA implementation. The attack was successful in recovering the private keys used to secure SSL connections. The main reason of the attack results is the lack of security controls and protections of DRAM allocation and access policies. This issue could be very critical for multi-tenant cloud platforms in which co-tenants have the immediate hardware access and control of the shared DRAM modules. For multi-tenant platform it has not been done yet.

\section{Countermeasures}
\label{Section:Countermeasures}
To prevents the attacks mentioned in section \ref{Section:Attacks} several methods has been proposed in the literature's \cite{cloud_security_1},  \cite{Jin2020SecuritySurvey}. We divided the methods into two major categories : 1. Tenant approach and 2.  Cloud Providers Approach

\subsection{Tenant Approach}

\subsubsection{Masking and Hiding} 
\label{masking_hiding}
Researchers have proposed many \textit{masking} and \textit{hiding} strategies for building attack side channel attack resilient circuits. To prevent attacker from obtaining circuits internal details, in masking strategy, the base circuit is transformed into a  secure large circuit by applying some cryptographic algorithm \cite{masking_ishai}. This large secure circuit is totally different from the base circuit in terms of internal LUT tables and boolean functions but functionally equivalent to base circuit. Even though the attacker can observe some internal details it cannot extract the whole circuit as changed scheme of implementation

Hiding strategies at aims reducing the Signal Noise Ration (SNR) at computational stages of the crypto cores which can be exploited by malicious attackers to leak information from the IP core. Generally, SNR reduction is implemented by either adding additional noise or reducing the strength of the signal power trace in computational phases \cite{hiding_1}, \cite{hiding_2}, \cite{hiding_3}.  
\subsubsection{IP Watermarking }

In a multi-tenant environment, tenants can integrate IP watermarking into their hardware design by introducing some unique data identifier into hardware design \cite{watermarking_1}. IP watermarking method could prevent potential IP counterfeits and reverse engineering threats \cite{watermarking_2}. By integrating the IP watermark, tenants can prove ownership of the hardware design and prevent illegal IP use from malicious cloud providers. Some watermark techniques could incur additional area overhead in routing paths. IP watermarking methods can be generally classified into
five groups: i) Constraint-based watermarking, ii) Digital
signal processing (DSP)-based watermarking, iii) Finite state
machine (FSM)-based watermarking, iv) Test structure-based
watermarking and v) Side channel-based watermarking \cite{watermarking_2}. The constraint-based watermarking method was introduced in research \cite{watermarking_constraint}, where the author's signature was mapped into a set of constraints that can hold some independent solution to complex NP-hard problems. FSM-based watermark adds additional FSM states at the behavioral level without altering the functionality of the hardware circuit. By modifying the state transition graphs, FSM watermarks can be embedded into the hardware design \cite{watermark_fsm}. Test structure-based watermarking would not fit in the context of multi-tenant cloud FPGA security. It depends on  IC supply chain testing methodology and focuses more on fabrication-based testing techniques. Digital Signal based watermarking could be introduced by slight modifications to the decibel  (dB) requirements of DSP filters without compromising their operation \cite{watermarking_dsp}.

\subsubsection{Obfuscation } 

Hardware obfuscation or camouflage is a unique approach to obfuscate tenant circuits from malicious attackers. In this method, the circuit's functionality is obscured to a functional equivalent code by secret cryptographic keys. The idea behind this attack is to modify the state transitions functions of a circuit so that it only generates desired output for correct input patterns. This same obfuscated circuit can generate incorrect functional behavior with provided wrong input patterns or keys \cite{counter_obfuscation_1}. By this method, the tenant can protect their Intellectual Property (IPs) from unauthorized manufacturing and cloning. Its also provide prevention for IP privacy issues \cite{counter_obfuscation_2}.  
 A more common practice of obfuscation in industry is widely adopted by HDL encryption \cite{counter_obfuscation_3}. The HDL source code is encrypted and obfuscated, and the IP vendor provides valid authentication keys to only registered licensed customers.



\subsection{Cloud Providers Aprroach }
\subsubsection{Bitstream Antivirus }

The encryption algorithms and key storages used by contemporary FPGAs from top suppliers are listed in Table \ref{tab:bitstream}.
The onboard volatile memory, which is powered by a coin-cell battery, is where the secret symmetric key is encrypted and kept.

\begin{table}[htbp]
	\caption{Bitstream encryption in modern FPGAs}
	\label{tab:bitstream}
	
	\centering
	\begin{adjustbox}{width=1\textwidth}
		\large
	
	\begin{tabular}{|c|c|c|c|c|c|}

		\toprule
Features	&	Virtex/Kintex Ultrascale+ &Zynq Ultrascale+&Aria-10 GX / SX  & Stratix-10 GX/SX & SmartFusion2 \\
		\midrule
Encryption Algorithm&		AES-GCM256 & AES-GCM 256 & AES-GCM 256 & AES-GCM 256 & AES-GCM 128/256 with ECDH   \\
Key Storage	&	BBRAM eFuse & 	BBRAM eFuse & 	BBRAM eFuse& AES-GCM 256 & AES-GCM 128/256 with ECDH   \\
	
		\bottomrule
	\end{tabular}

\end{adjustbox}
\end{table}

Tenants can encrypt their designs or region using bitstream encryption cores provided by FPGA vendors. But, in the cloud FPGA platform, the tenant hardware design is not delivered to the public cloud providers as a bitstream. Amazon AWS EC2 F1 only accepts its own customized synthesized netlist \cite{amazon}. This customization process includes reshaping reconfigurable regions, applying design rule checks, and verifying routing constraints for the tenant's IP blocks. Cloud providers can enforce strict design rule checking on the tenant's bitstream by bitstream checkers to prevent malicious side-channel attacks and fault-injection attacks \cite{baidu}. Bitstream checkers could identify malicious circuit structures, e.g., malicious ROs or combinational loops, and prevent bitstreams from being uploaded. Alternative designs, such as sequential RO circuits, can still avoid bitstream verification, and the standard technique used by AWS and other significant FPGA cloud providers is insufficient because it can only detect assaults using LUT-based ROs.  \cite{oscillators_sugawara}.  The first bitstream checker, a.k.a FPGA antivirus, was proposed by Gnad et al. \cite{bitstream_gnad} that checks for known negative patterns on the circuits. Similar to the antivirus concepts, this bitstream checker checks at signatures of malicious logic that might lead to electrical-level attacks. Instead of matching exact circuits, it intends to formulate the malicious attacks' fundamental properties and provide protection against fault injection and side-channel attacks. In the case of fault injection attacks, the checker extracts combinational cycles and evaluates a threshold number of cycles that can be allowed for the designer without being able to launch any crashes and fault attacks. As timing violations and TDC could incur side-channel leakage, this checker also prevents side-channel attacks by confirming the netlist has zero timing violation constraints.  While in the work \cite{bitstream_gnad}, LUT based ring oscillator designs are detected by scanning bitstream netlist. This work cannot prevents alternative oscillator design (e.g. glitch based oscillators)\cite{oscillators_sugawara}. 

FPGADEFENDER proposes an FPGA virus scanner that detects malicious non-ROs based designed by scanning bitstream covering the areas 
configuration blocks (CLBs), block ram memories (BRAMs), and signal processing blocks (DSPs) \cite{fpga_defender}. Beside ring oscillators, the most severe malicious FPGA threats are self-oscillators (SOs) based attacks. Unlike ROs circuits, self-oscillators (SOs) depend on the external clock and soft logic feedback and can hardly be detected by AWS and major FPGA vendors' bitstream checkers. FPGADEFENDER can detect these non-trivial self-oscillator circuits threats. In the work \cite{fpga_defender}, FPGADEFENDER designed and proposed a circuit using different self-oscillating circuits by means of combinatorial feedback loops and asynchronous flip-flop modes and scanned different non-RO-based oscillatory circuits. Although, one of the bottleneck of this work was to flag the non-malicious true random number generator (TRNG).

\subsubsection{Active Fence and side channel attack prevention}
Krautter et al. \cite{fence_krautter} propose implanting a fence between the tenant region and attacker circuit. This work aims to prevent side-channel attacks from neighboring attacker circuits by creating extra noise signals utilizing ROs chains. As a result, it reduces the circuit's total signal-to-noise(SNR) ratio at the electrical level. The work aims to provide electrical level defense against side-channel or voltage-based extraction attacks where logical isolation methods are ineffective. Both heuristic and arbitrary approaches were followed to place the ring oscillators between malicious circuits and tenant regions. In addition, besides the placement of ROs, two activation strategies were considered for efficient SNR reduction. The amount of activated ROs depends on a TRNG output in the first strategy. Secondly,  ROs are proportionally activated by the value of the voltage-fluctuation sensor. Besides this work, hiding and masking are popular countermeasures against side-channel attacks. However, as hiding and masking are more IP dependent, they are discussed in the Tenant Approach countermeasure (Section \ref{masking_hiding}) and should be provided under the tenant's responsibility.

\subsubsection{Access Control }

In an FPGA-accelerated multi-tenant platform, execution of an software application both depends on processor and FPGA fabric. Software part includes exploiting the advantage of built-in operating system or sequential programming, where FPGA fabric accelerates hardware computation.  To prevent unauthorized authentication and software access, trusted cloud provider entity can enforce security rules and different policies for its hardware region and attached hardware peripherals. Joel et. al. proposed a domain isolation based access control mechanism for multi-tenant cloud FPGAs \cite{mandebi2021domain}. The threat model in this work considers a malicious software on a virtual machine try to steal information by accessing the FPGA region assigned to another VM. The threat mitigation approach uses a hardware-software co-design method to enforce the access control policies to shared FPGA regions. They propose a central software-based policy server, hardware based policy enforcer named ``Hardware Mandatory Access Controller (HMAC)", a secured communication protocol between policy server and HMAC, and RO-TRNG based key generation for secure authentication. This work also does not consider the side channel based attacks on cloud based FPGA sharing. Access control based security rules checking in SoC-based embedded systems was discussed in work \cite{Festus2017}\cite{FestusFPL}.  

\subsubsection{Domain Isolation for cross-talk and covert channel attack prevention}

Crosstalks attacks have been discussed thoroughly in section \ref{crosstalk}.
Logical domain isolation amongst tenants is the most effective method to prevent crosstalk and cross-channel attacks.  Domain isolation solutions can be carried out either by physical isolation or logical isolation. Logical isolation rely on dividing FPGA region into software controlled secure region and enforcing security mechanism to provide secure execution of environment \cite{Google}, \cite{secpolicies}, \cite{xilinxtrust}, \cite{intelTEE}, \cite{sgx}, , \cite{TrustZoneARM}.



\begin{figure} [hbt]
\centering
\includegraphics[width=0.6\linewidth]{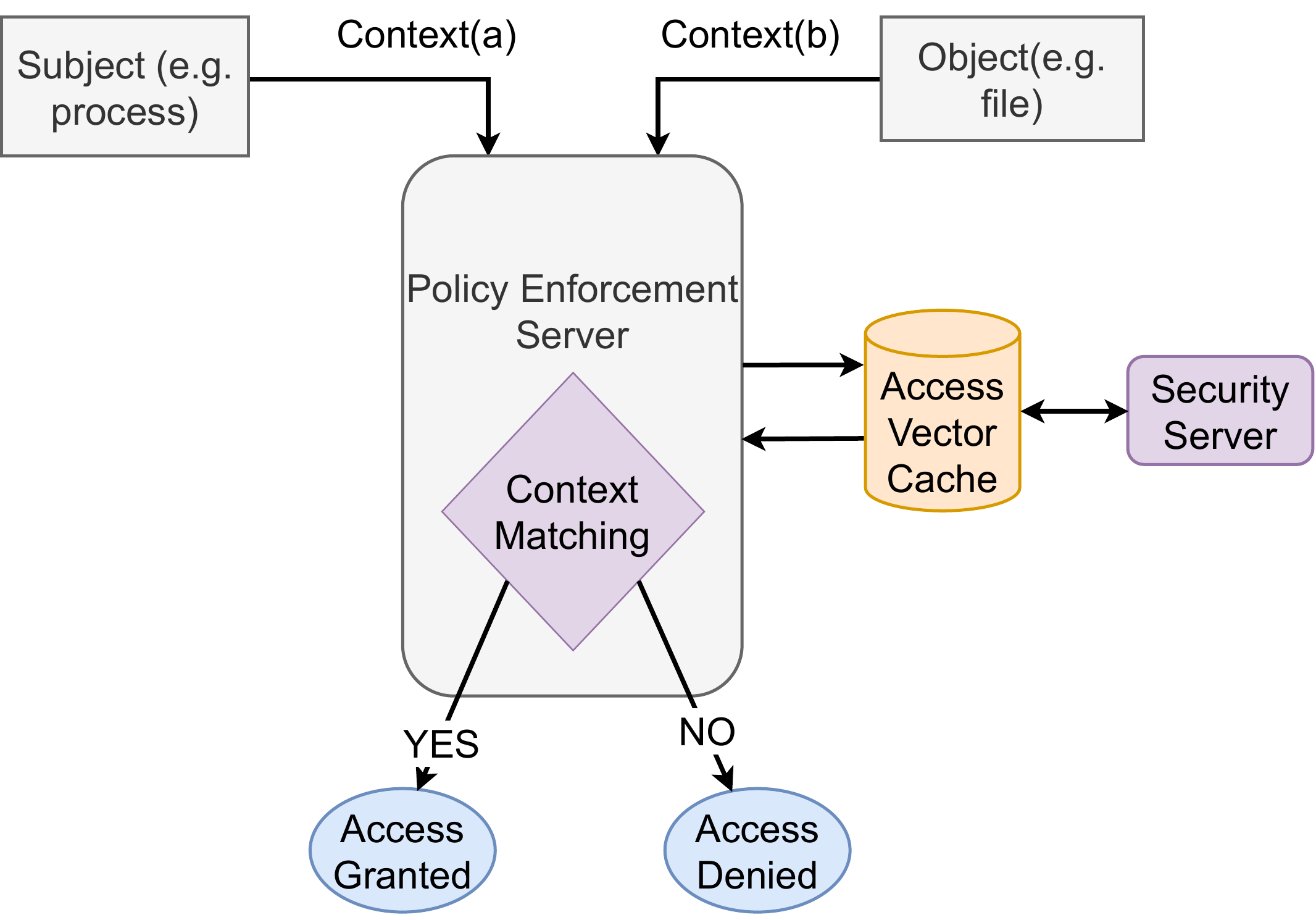}
\caption{  FLASK security architecture functionality \cite{MbongueHKAB18}.The policy server enforces IP context matching and compares the IP identity with the access vector cache that has been previously saved in secured storage server.   }
\label{fig:flask}
\end{figure}
In \cite{MetznerLB15, MbongueKB18, MbongueHKAB18} Joel et al. proposed logical hardware isolation technique (Fig \ref{fig:flask}) using FPGA virtualization without major performance loss. In context of multi-tenant cloud FPGA security, this work also investigates integration of access control along with hardware isolation. In SoC, for protecting IPs from tampering and crosstalk some hardware isolation strategies were presented in \cite{Festus15IPM, HategekimanaNB16} and further extended in \cite{festus, BobdaMWKK17, BobdaWKKN17} to shield hardware IPs using hardware sandboxes.  Some of the prior research on domain isolation in hardware has focused on isolating FPGA accelerators on SoC platform \cite{embeddedhuff}, \cite{Huffmiremem}, \cite{saha2020fpga}, \cite{proofcarrying}, \cite{proof}, \cite{proofcar}, \cite{referencem}, \cite{sapper}, \cite{intelreference}.

\begin{figure} [hbt]
\centering
\includegraphics[width=0.9\linewidth]{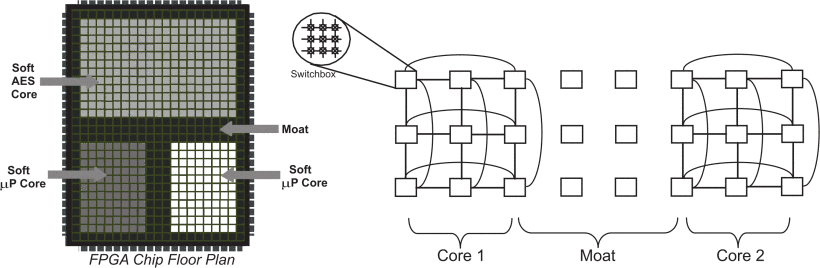}
\caption{  Physical hardware isolation technique by establishing moats and draw-bridge \cite{moats}}
\label{fig:moats}
\end{figure}
Huffmire et al. propose a physical hardware isolation technique (Moats and Draw-bridge) to isolate FPGA region for different tenants\cite{moats}. Physical hardware isolation is established by creating a block of wires("moats") around each core region. The core region can only communicate with other region via "draw-bridge" tunnel. This tunneling method is applied by disabling all the CLB switchboxes near isolated region.




\section{Virtualization Security Risks and  Future Challenges }

Using VMs and a VMM, a host system is virtualized in the multi-tenant cloud FPGA virtualization method.
This software-based virtualization layer's vulnerabilities could be used to compromise data. Unreliable IaaS has direct access and is able to go beyond the virtualization layer. As a result, the FPGA must take care of its own security and cannot rely on host computer software. 
In subsection \ref{subsection:OSSupport} we introduced existing operating system support for FPGA virtualization. Virtualization in FPGA is generally implemented using software or OS abstraction. In AmorphOS’s OS \cite{os_amorph}, space shared multi-tenancy is introduced by dedicating I/O and memory bandwidth to specific tenant (Morphlet). Although AmorphOS’s provides some protection similar to software OS , it does not provide explicit protection against remote side channel attacks \cite{os_amorph}.  

BORPH \cite{os_borph}, exploits and extends Linux Kernel for abstracting OS level instruction in the FPGA. Although, this proposed OS framework can enable OS in FPGAs, security primitives remains to be explored. Concepts such as secure system call, swapping , parallel file system access need to be refined. Scaling of BORPH model for increased demand without compromising security should be the priority in future research.  Due to the high overhead added by program image loading, Feniks OS mostly uses spatial sharing for multitasking rather than dynamic accelerator reloading (context switching) \cite{os_fenik}. Although, the literature claim that their custom accelerators templates will prevent collusion between OS and neighbour accelerators, the benefits was not demonstrated in the implementation. Morever, it doesnot provide any explicit protection or countermeasures against any established cloud FPGA adversaries.

\paragraph{\textbf{Future Challanges}}
FPGA acceleration in a cloud platform is a trending exploration, and we believe it will continue to rise in the forthcoming years. However, in the future years, more attacks are likely to be proposed on the multi-tenant platform. Current attacks described in section \ref{Section:Attacks} will be perhaps outdated and need more exclusive and comprehensive definitions.  

Cloud providers are more responsible and should come forward to provide strong security measures for tenants being protected from the malicious attackers. Existing logical and physical isolation methods \cite{embeddedhuff}, \cite{Huffmiremem}, \cite{saha2020fpga}, \cite{proof},  \cite{referencem},  \cite{intelreference} should be carefully extended and exploited in multi-tenant platform to establish secure isolation of tenants. Moreover, it should also explore some concrete and specific virtualization/isolation mechanisms for spatial sharing of the underlying FPGA fabric. Bitstream checkers provided by cloud providers should also be able to detect non-RO-based side-channel and fault injection attacks without compromising the tenant's design. Existing public cloud provider giants still lack deploying a runtime monitoring system for active defenses against real-time known attacks. One of the core challenges would be to provide secure access to reserved shared Shell region elements(memory, control logic, and PCIe buses) so that a tenant could only access its designated address and memory data. Any compromise of Shell region abstraction would hamper the isolation of these core system components and could potentially lead to launch row hammer attacks \cite{multi_rowhammer_1} and crosstalk attacks\cite{multi_cross_2}. Existing fault injection and side channel attacks detection mechanisms still hold a significant overhead in the overall circuit which need to be addressed. 

In multi-tenant platform, tenants still have to trust the major public cloud providers as they have to provide major architectural support for security. Beside, this tenants should focus on obfuscating their important critical design by hiding and masking strategies e.g. obfuscation, watermarking and noise reduction methods \cite{counter_obfuscation_2}, \cite{watermarking_2},\cite{hiding_2}. Although some of the hiding and masking and obfuscation strategies are not still extended in multi-tenant platform and currently being implemented in the SoC platforms, we believe they have great potentiality for increasing the security of the tenants IPs blocks and designs. Also, researchers are more likely to proposed various different low cost and low overhead side channel mitigation's countermeasures in the forthcoming years. Tenants could also use their cryptographic FPGA primitives for different security purposes such as IP authentication, privacy and integrity check, and establishing the root of trust. Most public cloud providers still restrict using RO-based combinational loops in the circuit, which is the fundamental block of many PUF and TRNG generators (e.g., Amazon EC2 F1 instance blocks RO-based design). Alternatively, for true random number generators(TRNGs), tenants can exploit some TRNG circuits that use meta-stability as a source of randomness \cite{countermeasure_trng_1}, \cite{countermeasure_trng_2}. For secure authentication, tenants can construct different FPGA-based physical unclonable functions (PUFs) circuits \cite{kawser}. Many cutting-edge PUF designs can be extended in the context of cloud FPGA and provide proof of trust and proof of execution fingerprint of the tenant.

\label{Section:Discussion}

\section{Conclusion}
FPGA accelerated cloud security research has evolved rapidly and should be exploited comprehensively in future years. Our survey discusses different recent works on multi-tenant cloud FPGA platform security threats and covers the proposed state-of-the-art defenses for existing security threats. We survey different FPGA deployment models in cloud platforms and aim to classify existing attacks. We also conclude our insights and provide guidelines for constructing and developing future security primitives in a multi-tenant context.

\subsubsection{Acknowledgments}

This work was funded by the National Science Foundation (NSF)
under Grant CNS 2007320

\bibliographystyle{ACM-Reference-Format}
\bibliography{references, sample-base,ref2,others,festus,fpt,cloud18,bobda}

\appendix

\end{document}